\documentclass{PoS}

\usepackage{graphicx}

\title{Progress with the LOFAR Imaging Pipeline}

\ShortTitle{LOFAR Imaging Pipeline}

\author{\speaker{George Heald}, John McKean, Roberto Pizzo, Ger van Diepen, Joris E. van Zwieten\\
        Netherlands Institute for Radio Astronomy (ASTRON), Postbus 2, 7990 AA Dwingeloo, the Netherlands\\
        E-mail: \email{heald@astron.nl}}

\author{Reinout J. van Weeren, David Rafferty, Sebastiaan van der Tol, Laura Birzan, Aleksandar Shulevski\\
	Leiden Observatory, Leiden University, Postbus 9513, Leiden, 2300 RA, the Netherlands}
	
\author{John Swinbank\\
	University of Amsterdam}
	
\author{Emanuela Orr\`u\\
	Radboud University Nijmegen, Heijendaalseweg 135, 6525 AJ Nijmegen, the Netherlands}
	
\author{Francesco De Gasperin\\
	Max Planck Institute for Astrophysics, Karl Schwarzschild Str. 1, 85741 Garching, Germany}

\author{Louise Ker\\
	SUPA, Institute for Astronomy, Royal Observatory of Edinburgh, Blackford Hill, Edinburgh EH9 3HJ, Scotland}

\author{Annalisa Bonafede, Giulia Macario\\
	INAF - Istituto di Radioastronomia, via Gobetti 101, 40129 Bologna, Italy}

\author{Chiara Ferrari\\
	UNS, CNRS UMR 6202 Cassiop\'ee, Observatoire de la C\^ote d'Azur, Nice, France}
	
\author{on behalf of the LOFAR Collaboration}

\abstract{One of the science drivers of the new Low Frequency Array (LOFAR) is large-area surveys of the low-frequency radio sky. Realizing this goal requires automated processing of the interferometric data, such that fully calibrated images are produced by the system during survey operations. The LOFAR Imaging Pipeline is the tool intended for this purpose, and is now undergoing significant commissioning work. The pipeline is now functional as an automated processing chain. Here we present several recent LOFAR images that have been produced during the still ongoing commissioning period. These early LOFAR images are representative of some of the science goals of the commissioning team members.}

\FullConference{ISKAF2010 Science Meeting \\
		 June 10 -14 2010\\
		 Assen, the Netherlands}

\begin{document}

\section{Introduction}

The Low Frequency Array (LOFAR; \cite{falcke_etal_2007}) is a unique radio telescope facility, and is a pathfinder for the Square Kilometre Array (SKA). Based on simple and inexpensive hardware components in the field\footnote{The current status of the buildup of LOFAR stations can be seen as a Google Maps overlay at {\tt http://www.astron.nl/$\sim$heald/lofarStatusMap.html}}, LOFAR's real strength as a cutting-edge observatory is in the backend processing that takes the raw data from the dipoles and generates (among other things) images of the sky. The huge data volume of a typical LOFAR observation (see \S\,\ref{subsection:dppp}), and the scope of the planned LOFAR coverage of the northern sky \cite{rottgering_tv}, together require that most image processing is done offline in a fully automated process. The focus of this contribution is on the Imaging Pipeline, which is being developed to achieve this automation. The Imaging Pipeline is an adaptation of the LOFAR Transients Pipeline \cite{swinbank_tv}.

\begin{figure}\label{figure:pipeline}
\begin{center}
\includegraphics[width=0.8\textwidth]{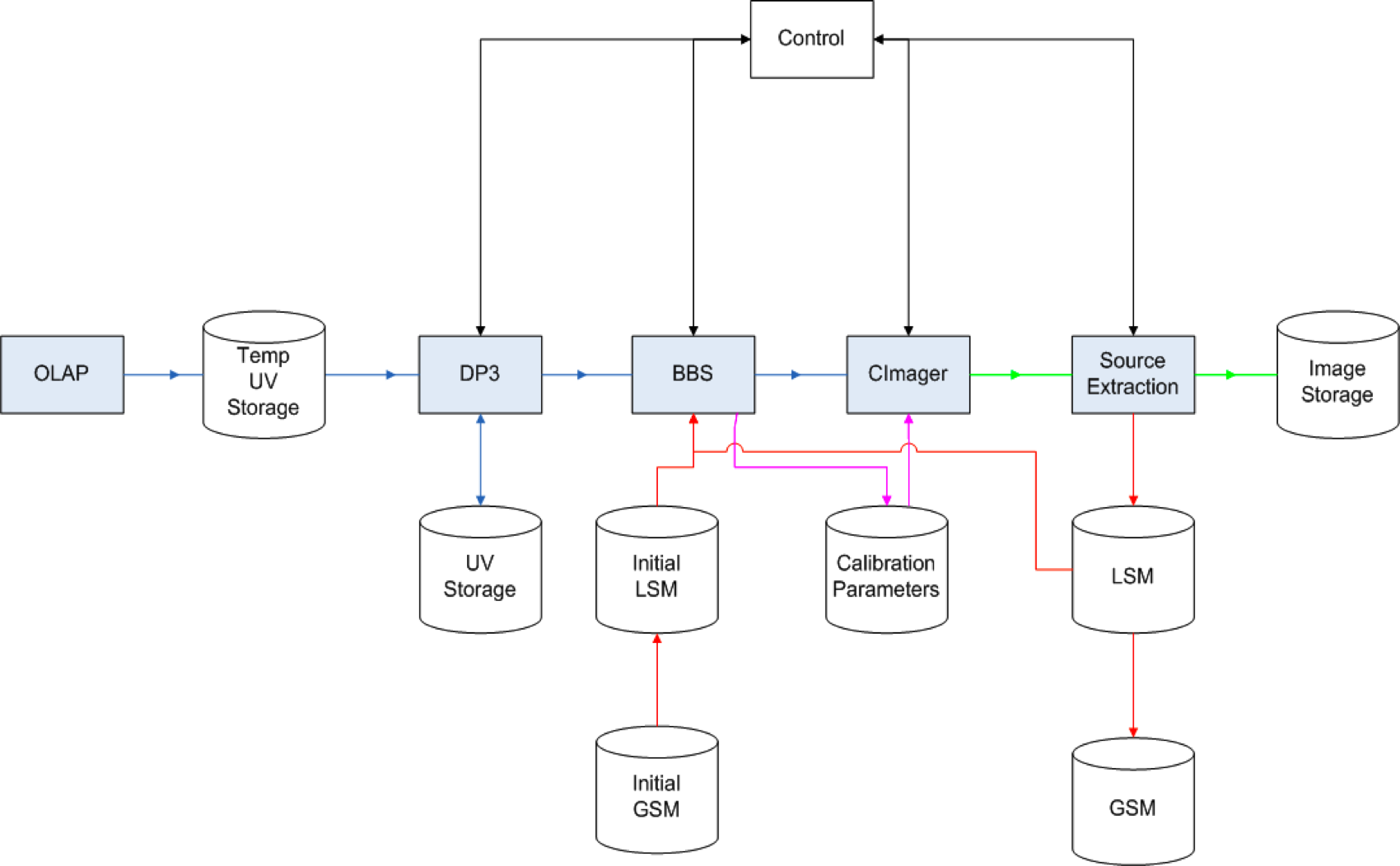}
\caption{The LOFAR Imaging Pipeline presented in schematic form, courtesy of R.~Nijboer (ASTRON). See the text for a description of the various software components and the data path.}
\end{center}
\end{figure}

The Imaging Pipeline is shown schematically in Figure \ref{figure:pipeline}. A detailed description of the individual components of the pipeline is provided in Section \ref{section:calibration}, but first we overview the full process. Following the data path from the left, visibility data are stored in Measurement Sets at the IBM Blue Gene/P correlator in Groningen (referred to as Offline Processing, or OLAP), and recorded to storage nodes in the current LOFAR offline processing cluster. The first data processing step is to flag the data in time and frequency, and optionally to compress the data in time, frequency, or both (in the Figure: Default Pre-Processing Pipeline, or DPPP). In future this stage of the processing will also include subtraction of the contributions of the brightest sources in the sky from the visibilities. Next, the calibration steps begin. Using the BlackBoard Selfcal (BBS) system that has been developed for LOFAR, a local sky model (LSM) is generated from a Global Sky Model (GSM)\footnote{Development of the initial GSM is the primary goal of the LOFAR Million Source Shallow Survey (MSSS), which is expected to begin in late 2010.} that is stored in a database. Calibration of the complex station gains is achieved using this LSM. At this stage, an additional flagging operation (not shown in the Figure) is performed with DPPP in order to clip any remaining RFI or bad data. Next, the calibrated data are imaged using either the CImager (which is under development for the ASKAP telescope \cite{johnston_etal_2008}), or more typically at present, the imager provided as part of the CASA package. In the commissioning process so far, we have deconvolved the resulting images using the standard Clark CLEAN algorithm, and updated the local sky model based on the resulting CLEAN components. We are currently working on a more sophisticated approach using source finding and characterization software to generate the updated local sky model. One or more `major cycle' loops of calibration, flagging, imaging, and LSM updates are performed. At the end of the process, the final LSM will be used to update the GSM, and final image products will be available.

The pipeline is currently implemented in python. The reader is referred to \cite{swinbank_tv} for a more complete description of the framework itself. The existing pipeline can be run, with no user intervention, on a full dataset to produce Stokes-I images after a single major cycle.

In the remainder of this contribution, we describe the main individual components of the Imaging Pipeline (DPPP and BBS) and illustrate actual LOFAR interferometric data at various stages of processing (\S\,\ref{section:calibration}). Recent LOFAR imaging results are described and illustrated in \S\,\ref{section:recentresults}. We conclude by listing some planned future work on the LOFAR Imaging Pipeline (\S\,\ref{section:futurework}).

\section{Calibration}\label{section:calibration}

\subsection{DPPP: flagging and data compression}\label{subsection:dppp}

The initial processing stage, DPPP, performs flagging and data compression. The default mode of DPPP operation makes use of median filtering in both the frequency and time axes to identify outlier data. A set of median window sizes, appropriate for LOFAR's RFI environment, have been defined through exhaustive testing. Using these settings, DPPP successfully identifies and flags RFI which is either narrowband or broadband, and either short- or long-duration. Optionally, we also make use of the flagger developed by Offringa \cite{offringa_etal_2010}.

One of the concerns about the construction of LOFAR in the Netherlands has been the high density of RFI in the local low-frequency radio spectrum. Indeed, the raw data show clearly that the spectrum contains numerous contributions from interfering sources. With this in mind, LOFAR was designed to provide extremely high frequency and time resolution during normal interferometric operations. The default frequency resolution is 763 Hz (each 195 kHz subband is made up of 256 channels), and typical integration times are either 1 second (in the low band LBA, $30-80$ MHz) or 3 seconds (in the high band HBA, $120-240$ MHz). The first flagging operation takes place on the full resolution data.

\begin{figure}
\begin{center}
\includegraphics[width=0.5\textwidth,angle=270]{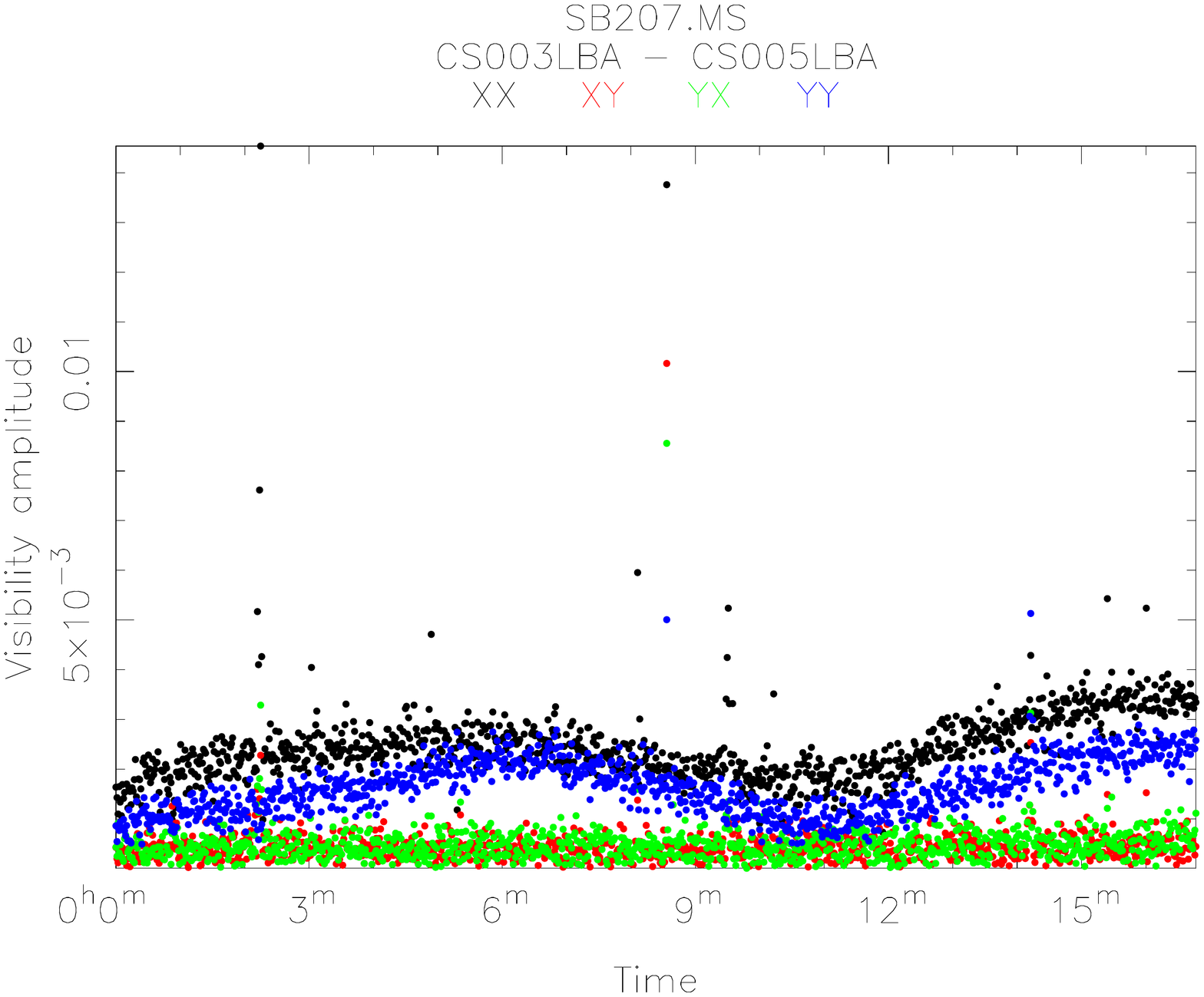}\\\includegraphics[width=0.5\textwidth,angle=270]{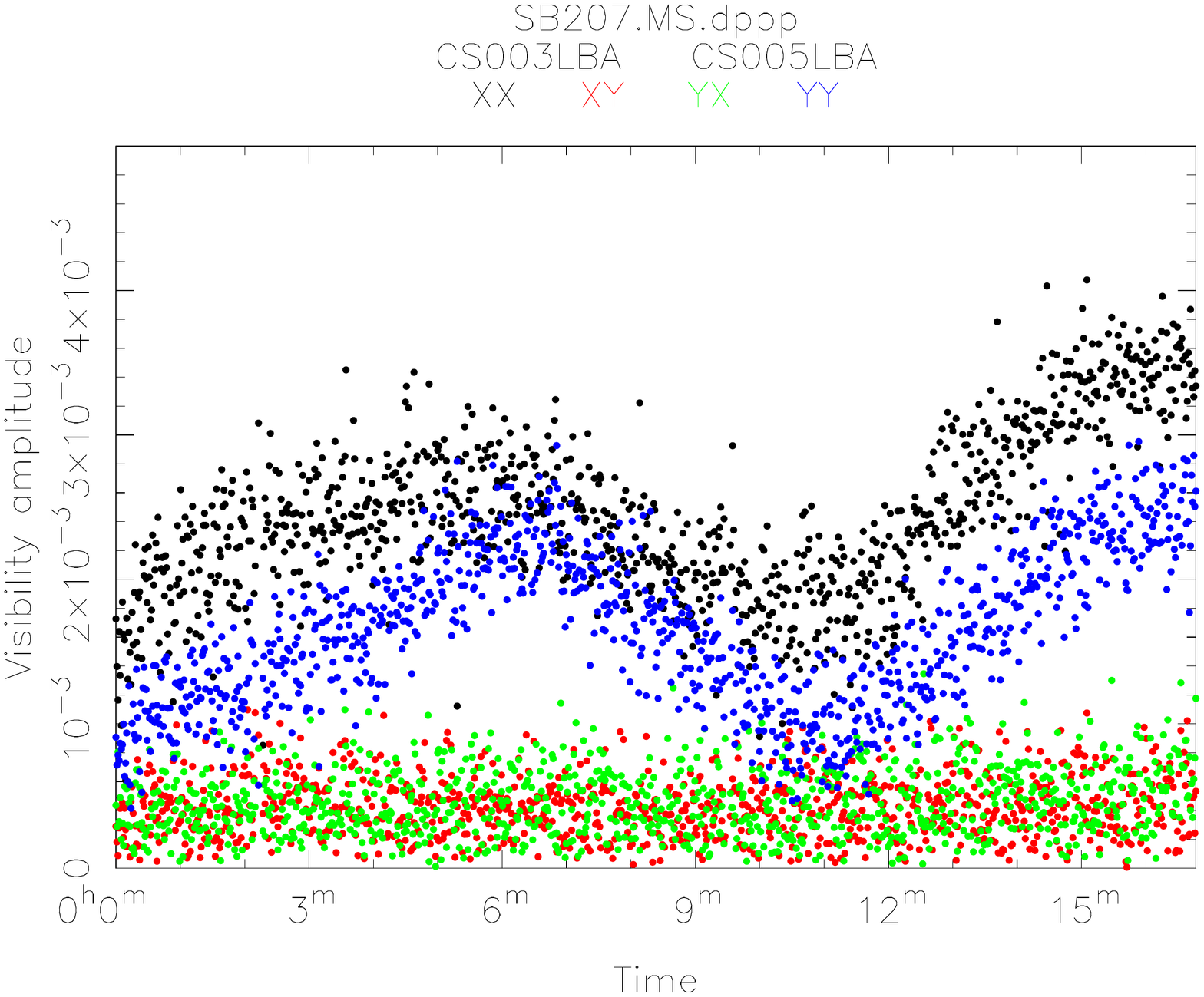}
\caption{LOFAR data before and after flagging with DPPP. The plots show visibility amplitudes on a short (approximately 250~m) baseline in the LOFAR core. The data in the top panel have been averaged in frequency (bandwidth 0.2~MHz) for presentation purposes. The short duration spikes in the visibility amplitudes are due to the presence of intermittent narrow-band RFI. After flagging, the spikes are seen to have been removed (bottom panel). In both panels, the scale of the vertical axis corresponds to the full range of visibility amplitudes in the time range displayed.}
\label{figure:flagging}
\end{center}
\end{figure}

To illustrate the appearance of a typical LOFAR observation before and after flagging, we present in Figure \ref{figure:flagging} a dataset before and after DPPP. Some regions of the spectrum have a much higher RFI occupancy than others, and these will be identified and avoided for future observations.

Typically, after the first flagging operation, we compress by a large factor in frequency ($16-256$, depending on the frequency band and the characteristics of the field) before proceeding to later flagging operations and data calibration. This averaging is done to reduce the data volume so that later processing is not prohibitively expensive. A typical LOFAR observation with a subset ($\sim70\%$) of the currently existing array (17 LBA stations, 248 subbands, 1 second integration time, 10 minute duration) consumes about 184 GB of disk space. Note that high band observations may be considerably bigger -- core HBA stations are split into two substations, and these may (optionally) be correlated as separate stations. Thus, the data volume of a core-station HBA observation can be about 4 times larger than a core-station LBA observation, all else being equal.

After calibration (\S\,\ref{subsection:bbs}), another short flagging and compression run is performed. The flagging step uses a large median window in time, and is designed to catch spikes caused by residual RFI or bad data. Finally, the data are often averaged in time by a factor of typically $3-10$ before the imaging step takes place.

\subsection{BBS: station gain calibration}\label{subsection:bbs}

The calibration step is performed using BBS. This calibration package is based on the Hamaker-Bregman-Sault Measurement Equation (ME) \cite{hamaker_etal_1996}, which expresses the instrumental response to incoming electromagnetic radiation within the framework of a matrix formalism. Here, the various instrumental effects are identified, their effect on the signal is characterized in full polarization, and are quantified and parameterized as separate Jones matrices. Each of these terms may depend on different dimensions: frequency (e.g. the bandpass); time (e.g. the station gains); or direction (e.g. the station beam). Because it is based on the general form of the ME, BBS can natively handle difficult problems such as direction dependent effects and full polarization calibration.

A critical input to BBS is the sky model which is used to predict the visibilities. Earlier in the commissioning process, one input to the BBS stage of the pipeline has been a hand-crafted listing of the brightest sources in the field of interest. We are currently testing an addition to the pipeline which automatically constructs an initial sky model based on cataloged values from the VLA Low-frequency Sky Survey (VLSS; \cite{cohen_etal_2007}). Note that this should be considered the ``Mark-0'' LOFAR GSM; the ``Mark-1'' LOFAR GSM will be generated by the upcoming MSSS, as mentioned above. MSSS will provide a higher areal density of sources than the VLSS catalog, and more importantly will include spectral information from 30 MHz up to about 180 MHz.

\begin{figure}
\begin{center}
\includegraphics[width=0.5\textwidth]{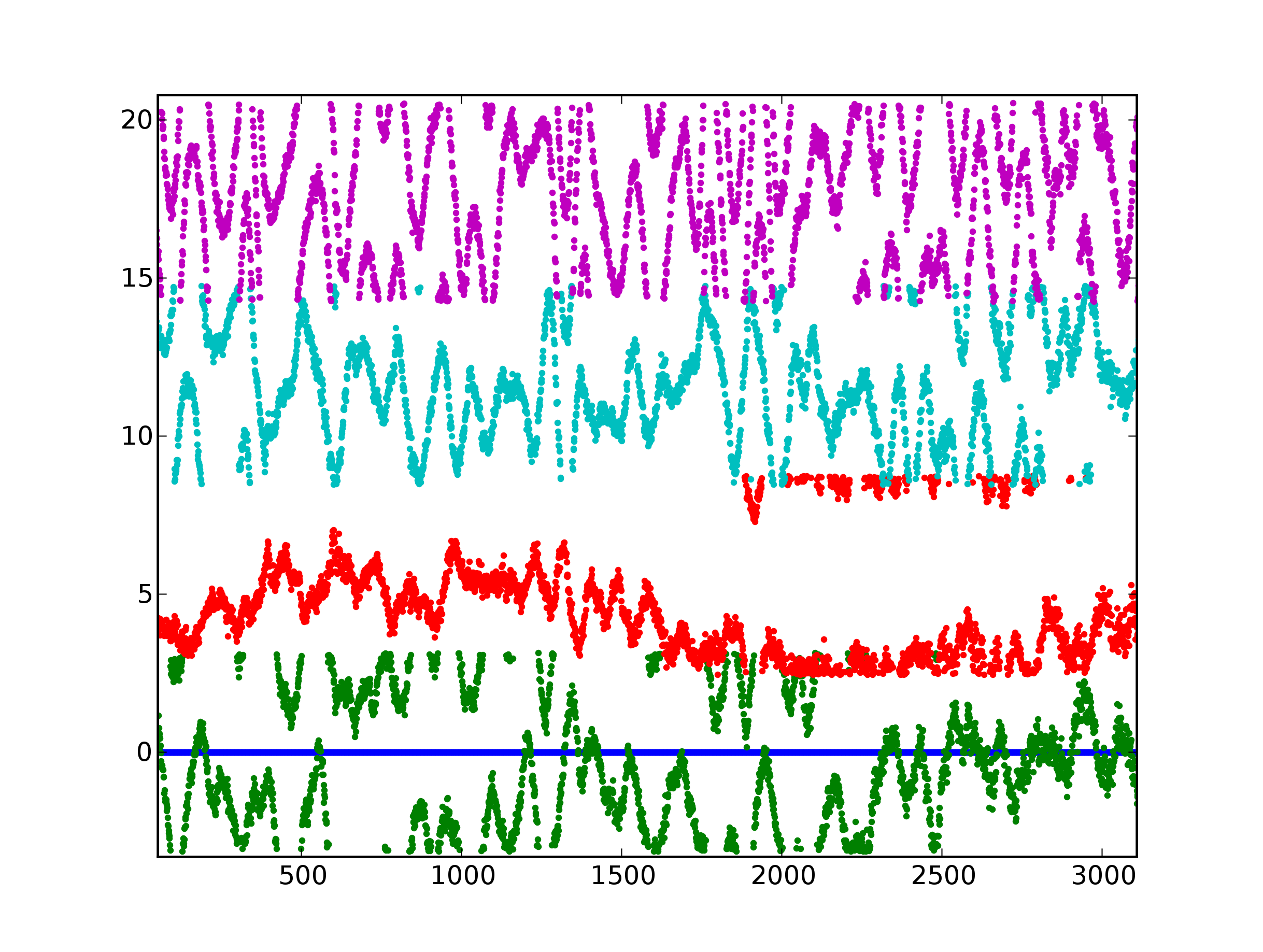}\includegraphics[width=0.5\textwidth]{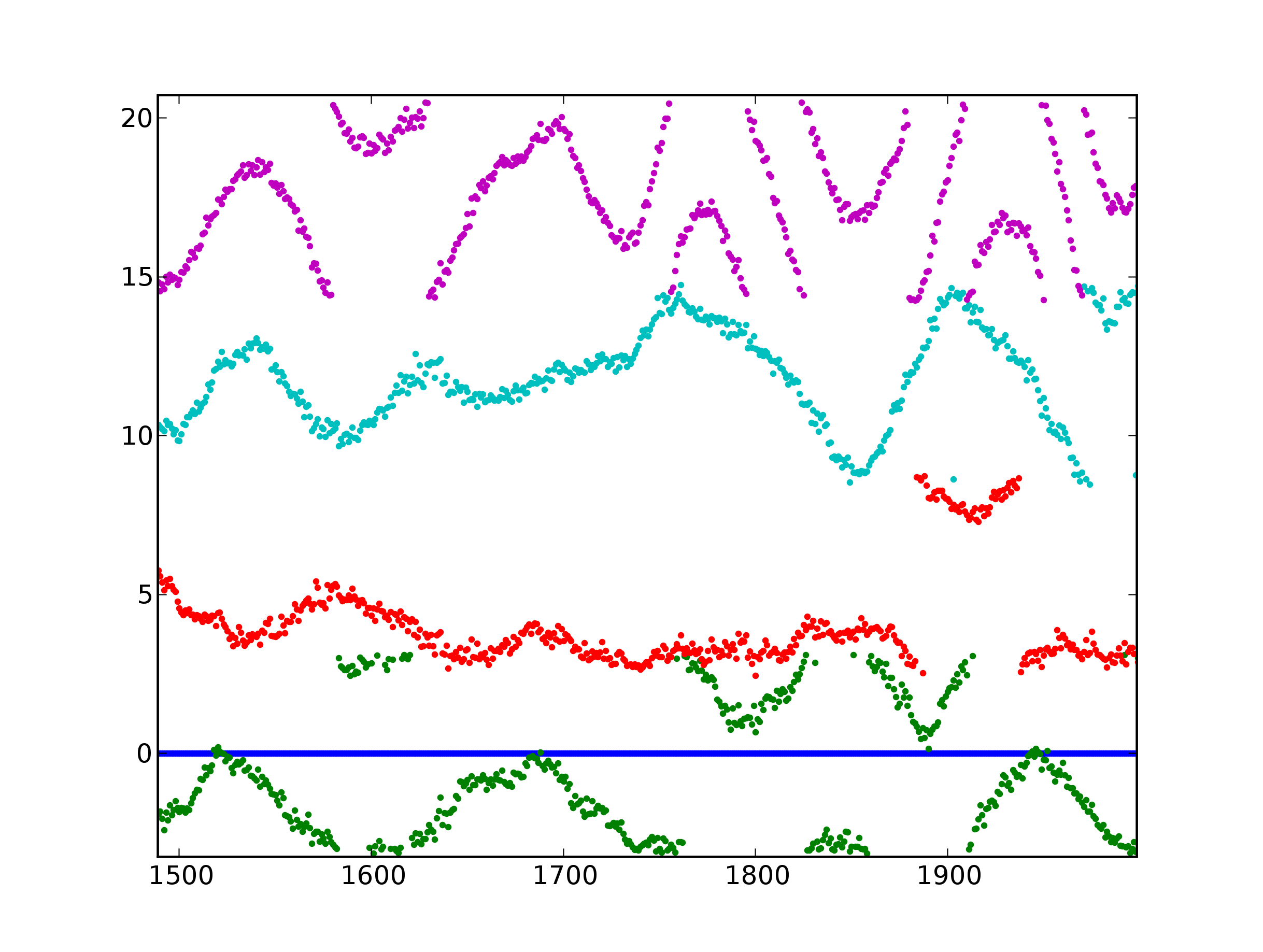}
\caption{Station gain phases determined by BBS. The blue line corresponds to the reference station. The vertical axis shows the phase in radians, plus an arbitrary offset for display purposes. The horizontal axis is the time in seconds. The right-hand panel shows a short segment of the phases in the left-hand panel, zoomed in to highlight the stability of the gain solutions.}
\label{figure:bbsgains}
\end{center}
\end{figure}

In Figure \ref{figure:bbsgains}, we illustrate station gain phases solved for by BBS. Gain solutions are shown for an early observation of 3C196 taken during August 2009. Five LBA stations (one core station, acting as the reference station, and four remote stations) were involved in this observation. A separate gain solution was obtained for every 1-second integration in the observation. In order of increasing distance from the core station CS302, gain solutions (corresponding to the Y-dipole) are shown for RS503 (red), RS106 (cyan), RS307 (green), and RS208 (purple). The gain phase variations relative to the core station are more rapid for more distant stations, as expected. At the farthest station used in this dataset, located at a distance of about 26 km from the reference station, variations corresponding to a full turn of phase within periods as short as 40 seconds are seen to have been corrected for by BBS. It should be noted that 3C196 is a very bright source which dominates the flux in this observation. In other fields, larger solution intervals, or averaging of multiple subbands, will be required to obtain high quality station gains.

\section{Recent Results}\label{section:recentresults}

During May 2010, a campaign of imaging commissioning, with an emphasis on complicated fields containing extended sources, was started in earnest. In this section we summarize the intermediate results of that campaign.

\subsection{Cygnus A}

The enigmatic radio galaxy Cygnus A is one of the brightest sources in  
the LOFAR sky. It is one of the objects that makes up the `A-team'; a  
small number of very strong radio sources that can potentially  
contaminate every LOFAR observation. To reduce the effect of these  
sources on our science observations, we must image each member of the  
A-team over the LOFAR observing band and remove them from the {\it uv}-data of our target fields. The large luminosity and small distance to  
Cygnus A also makes this source interesting for studying the  
properties of AGN, for example, to investigate feedback processes. Its  
complex double structure also makes for an excellent commissioning  
target to test the LOFAR system.

\begin{figure}
\begin{center}
\includegraphics[width=\textwidth]{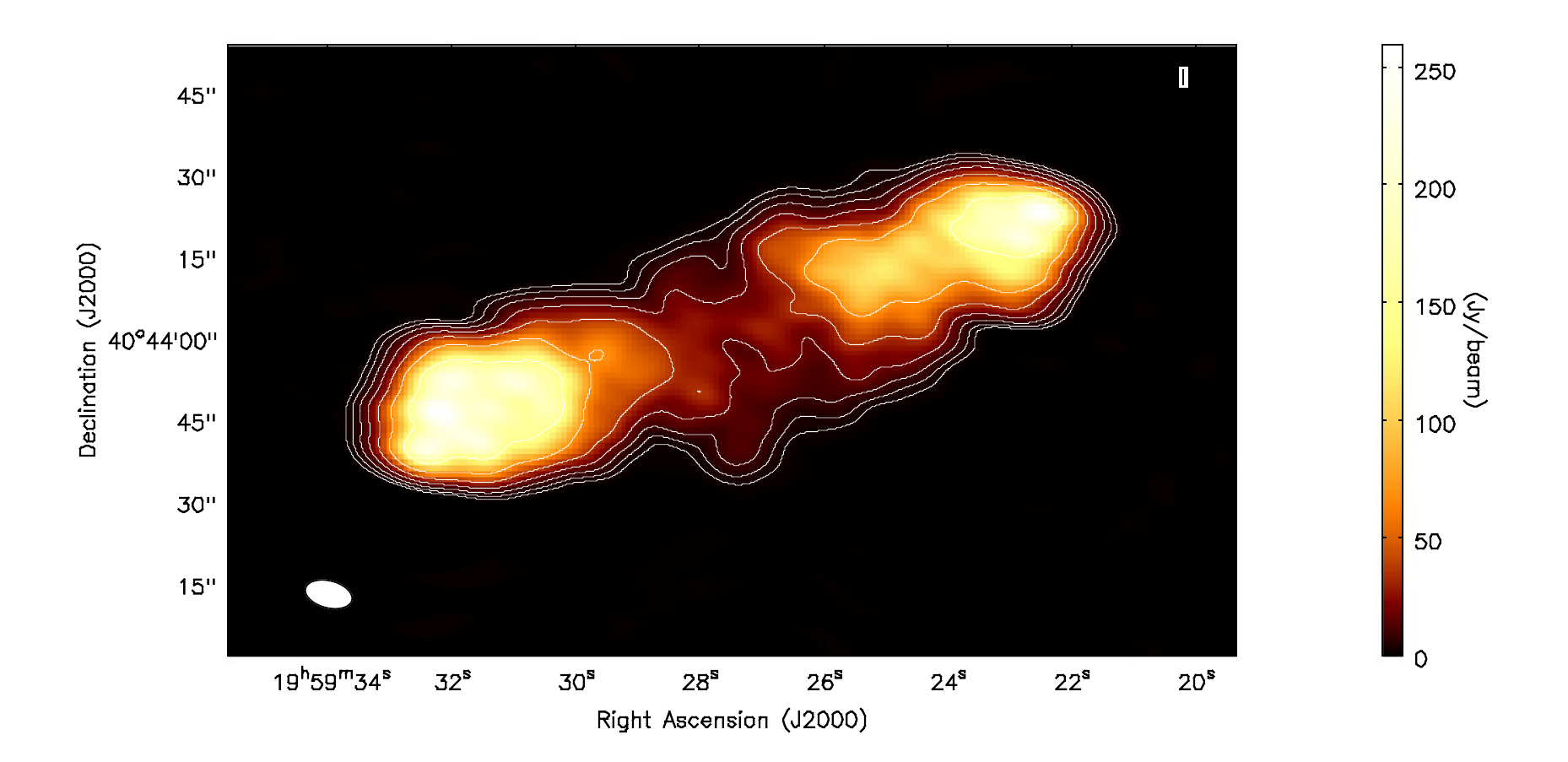}
\caption{LOFAR image of Cygnus A, as described in the text. The contour levels begin at 2.1 Jy beam$^{-1}$ and increase by powers of two.}
\label{figure:cyga}
\end{center}
\end{figure}

Cygnus A was observed for 6 hours in the LOFAR high band (210--240  
MHz) on 2010 May 30. The array consisted of 12 core stations, 4 remote 
stations and 1 international station. However, after data editing and  
calibration (within CASA), only 8 core stations and 4 remote stations  
were used for imaging. A single sub-band (0.2 MHz bandwidth) image of  
Cygnus A at 239 MHz is shown in Figure \ref{figure:cyga}. This image  
is uniformly weighted and has an elliptical restoring beam with a FWHM  
of $9 \times 5$~arcsec at a position angle of 79~degrees. The rms  
noise in the image is 0.3 Jy~beam$^{-1}$, giving a dynamic range of $ 
\sim$1000. The limiting factor in the dynamic range is the cleaning.  
The image shows the expected double lobe structure that has been  
observed from Cygnus A at other frequencies (for example, note the similarity to the 330 MHz image shown by \cite{kassim_etal_1993}). We also see complex  
structure within the lobes and in the space between them. The next  
step is to make an image with the full bandwidth and investigate the  
broad-band spectral properties of the features we observe.

\subsection{Abell 2256}

Galaxy clusters are the largest gravitationally bound objects in the universe. These clusters also contain a large amount of hot gas which emits at X-ray wavelengths. Galaxy clusters grow by mergers with other clusters and galaxy groups. During a merger event large shock waves are created which heat the gas in the cluster. It is thought that within these shocks particles are accelerated to highly relativistic energies \cite{drury_1983}. In the presence of a magnetic field these particles will emit synchrotron radiation that is visible at radio wavelengths. It has been suggested that such synchrotron emitting regions can be identified with so-called radio relics - diffuse, steep spectrum, radio sources found in some clusters \cite{ensslin_etal_1998}. Besides radio relics, some clusters also contain a diffuse radio halo. Radio halos could be the result of turbulence in the hot gas, caused by a cluster merger event.

\begin{figure}
\begin{center}
\includegraphics[width=\textwidth]{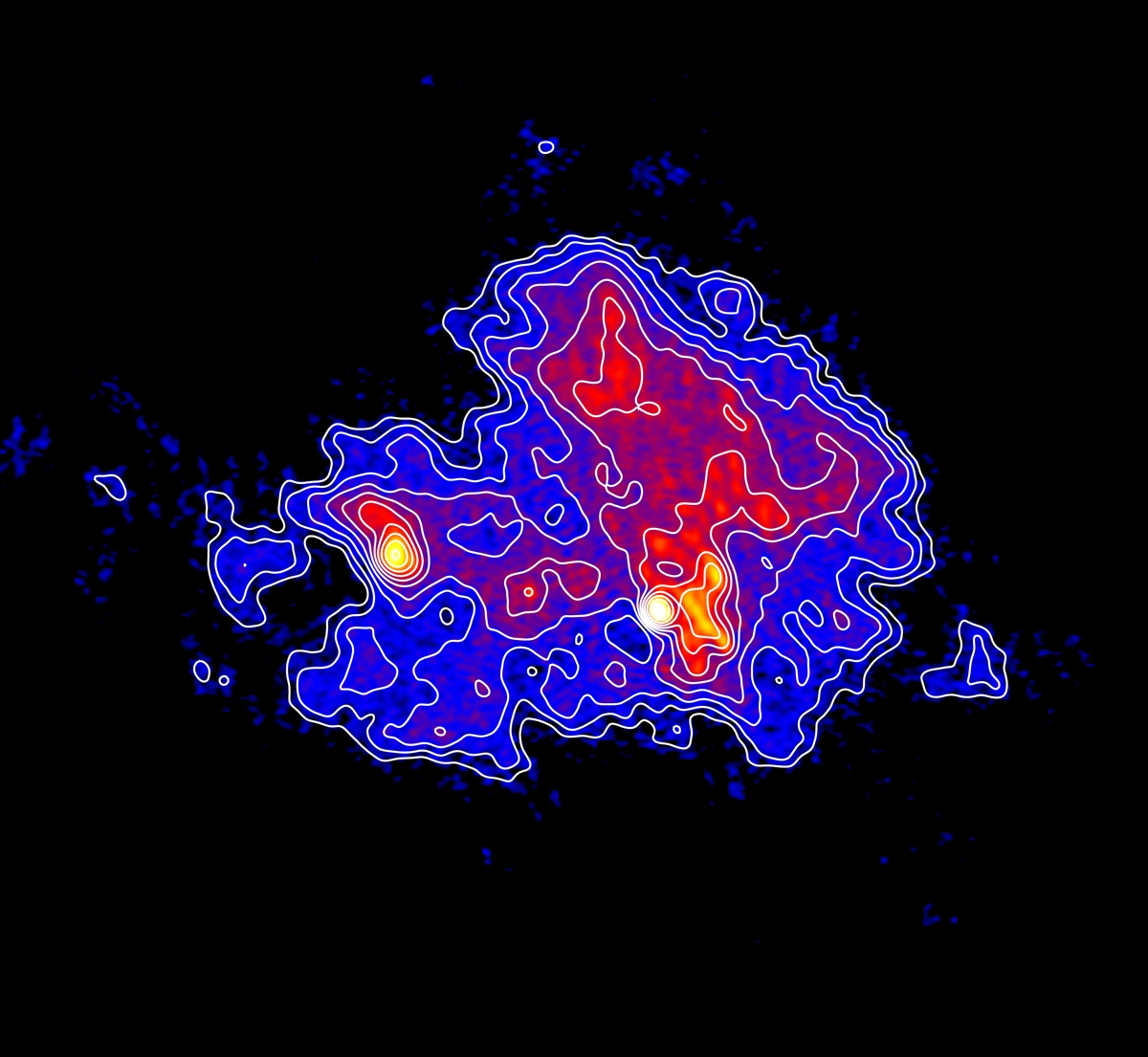}
\caption{LOFAR image of Abell 2256, as described in the text. The contour levels begin at 22 mJy beam$^{-1}$ and increase in increments of 25\%.}
\label{figure:a2256}
\end{center}
\end{figure}

A textbook example of a cluster hosting a radio relic and halo is Abell 2256 (e.g. \cite{clarke_ensslin_2006,brentjens_2008,intema_2009,vanweeren_etal_2009,kale_dwarakanath_2010}). This cluster was observed with LOFAR in the HBA band ($115-165$ MHz) in May 2010 for about 8 hours. The data were taken with 10 core stations and 5 remote stations (the core stations were split resulting in a total of 25 stations). The image (see Figure \ref{figure:a2256}) was made using 18 subbands covering a total of 4 MHz of bandwidth around 135 MHz. The resolution of the image is $31^{\prime\prime}\times19^{\prime\prime}$ and the noise is $\approx5\,\mathrm{mJy\,beam^{-1}}$, making it one of the deeper images of this cluster at low frequencies. The next step will be to combine all 240 available subbands to create an even deeper image. The yellow (brighter) regions in the image are associated with several disturbed radio galaxies. The large-scale emission in red and blue is mostly from the radio relic, although some additional faint emission in the south of the cluster (bottom of the image) comes from the radio halo.

\subsection{Messier 51}

Radio continuum emission from nearby galaxies is produced by both thermal (free-free) and nonthermal (synchrotron) processes. At low frequencies, the thermal contribution is minimal, and the spectrum is dominated by synchrotron emission. Because the steepness of the synchrotron spectrum is related to the energy spectrum of the emitting cosmic ray (CR) electrons, LOFAR observations are expected to provide crucial information about CR energetics and transport in galaxies. The CR energy spectrum should become progressively steeper in the outer parts of galaxies, as the CRs propagate far from the regions of star formation and suffer synchrotron energy losses. Low-frequency radio emission should thus reveal new information about the nonthermal component of galaxies by probing to much larger radii than has been available so far.

\begin{figure}
\begin{center}
\includegraphics[height=0.45\textwidth]{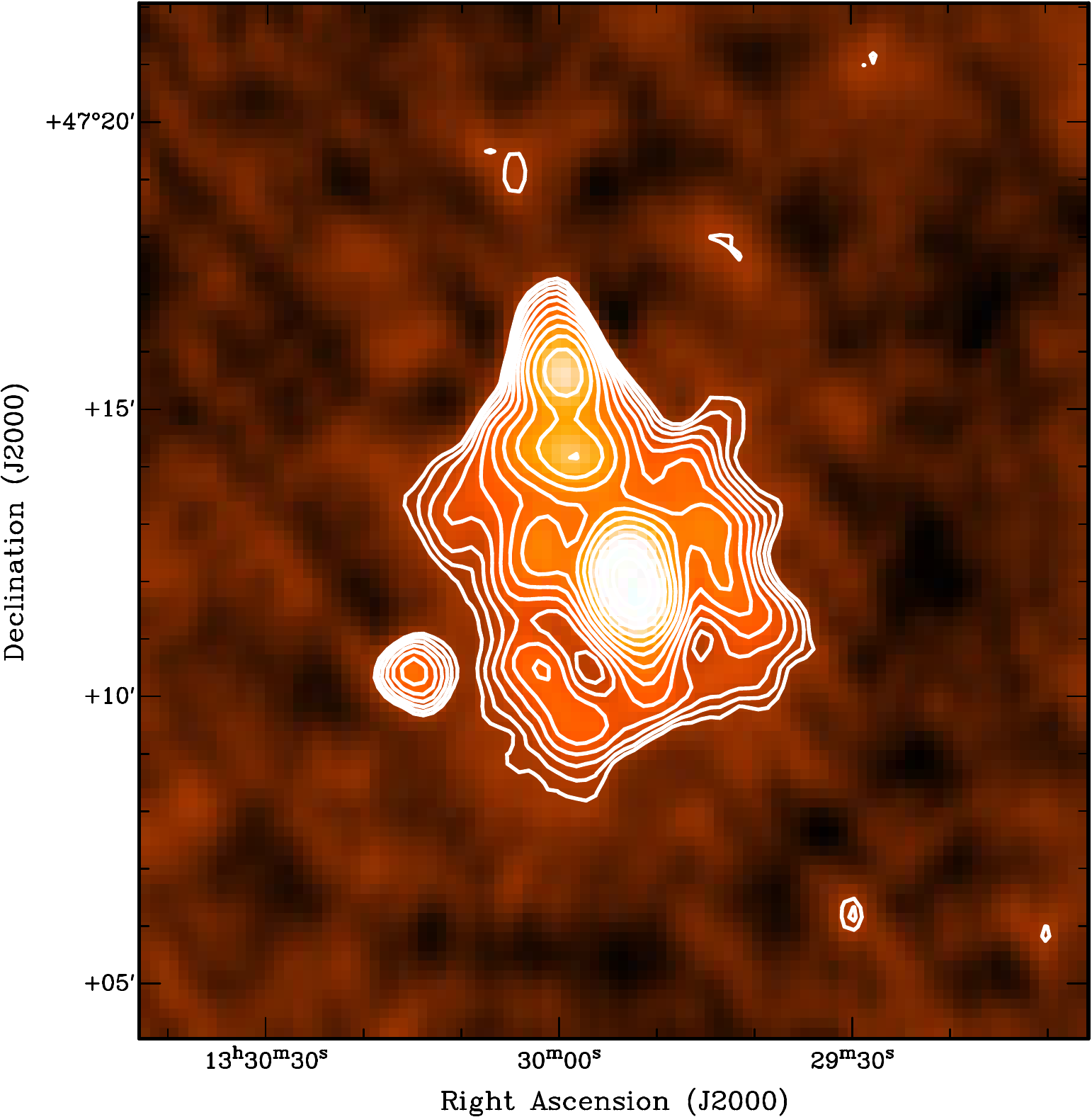}\includegraphics[height=0.45\textwidth]{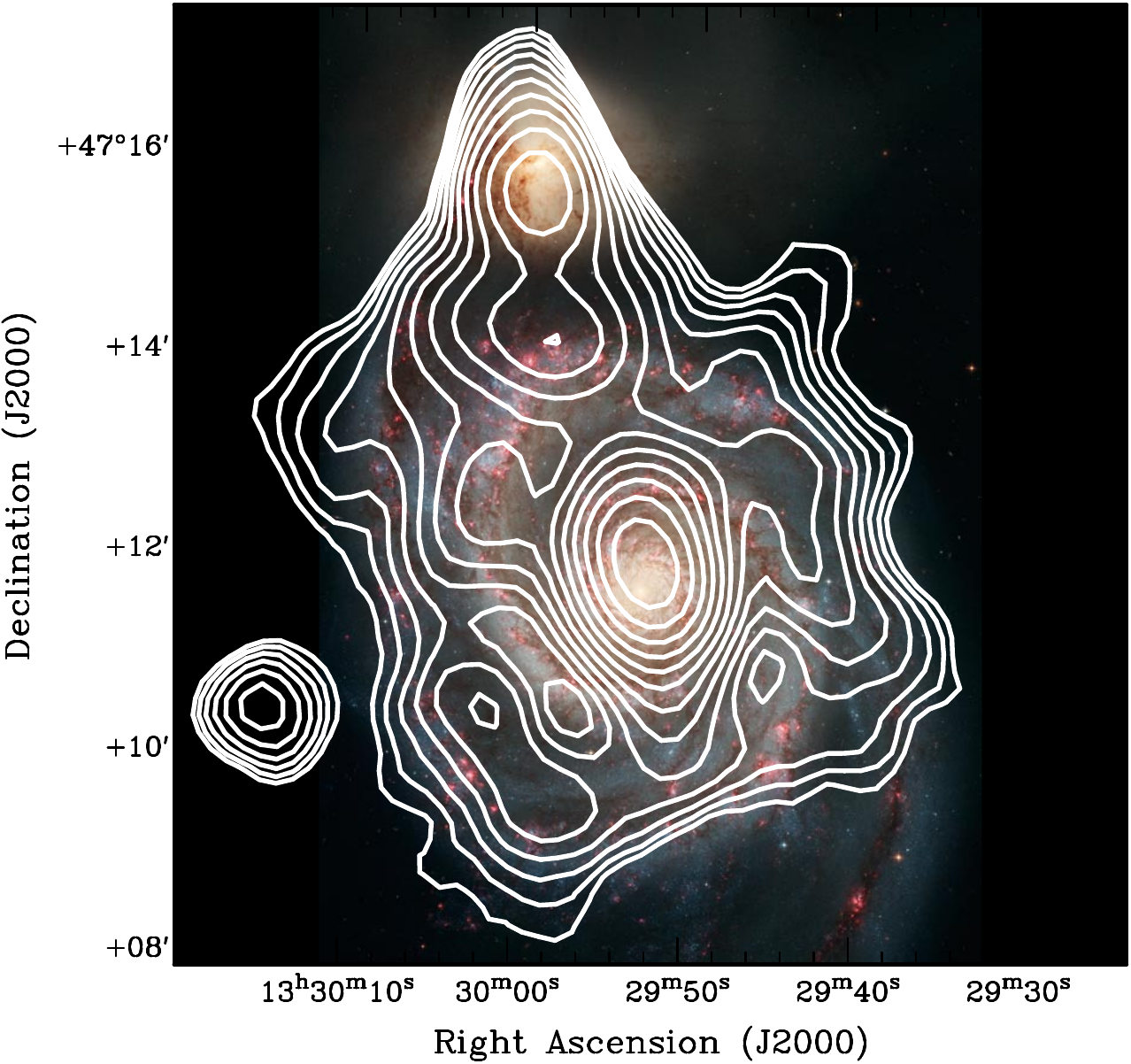}
\caption{LOFAR image of M51, as described in the text. The contour levels in both frames start at 15 mJy beam$^{-1}$ and increase in 25\% increments. The optical image in the right-hand panel is the Hubble Heritage image of M51.}
\label{figure:m51}
\end{center}
\end{figure}

The nearby spiral galaxy M51 (NGC~5194) was observed in the high band, for a total integration time of 4 hours. Twelve core stations and four remote stations were included in the observation. The core stations were used in their split mode, so the effective number of stations in the observation was 28. The total bandwidth of the observation was only 10 MHz, distributed throughout a frequency window spanning the range $120-155\,\mathrm{MHz}$. Unlike many other fields used for LOFAR commissioning, this field does not contain a single source that dominates the flux. An initial sky model was derived from an existing WSRT image at 140~MHz, provided courtesy of G. de Bruyn. Following a direction-independent calibration, a widefield image was produced with a low resolution of about 2~arcminutes. The rms noise in this image was 3 mJy beam$^{-1}$, and led to a dynamic range in the image of 550. This dynamic range was strongly limited by deconvolution errors corresponding to a bright off-axis source, 3C289. That source is 3.6 degrees from M51, and is actually the brightest source in the field by a small margin. Future work will involve direction-dependent calibration, which will be required to remove the strong artifacts and approach a more typical dynamic range. An image zoomed in on the vicinity of M51 is shown in Figure \ref{figure:m51}. Because the outer $uv$ coverage of the existing LOFAR array is rather sparse, care is required to optimize the balance between resolution and sensitivity when imaging an extended source like M51. The image shown here has an rms noise level of about $6\,\mathrm{mJy\,beam^{-1}}$ and a synthesized beam size of $59^{\prime\prime}\times53^{\prime\prime}$.

\section{Conclusions and Future Work}\label{section:futurework}

The imaging results shown here and elsewhere in these proceedings (see, e.g., \cite{debruyn_tv}) illustrate that the still growing LOFAR array works well for sub-arcminute imaging at low radio frequencies. Typical observations yield images with rms noise levels of order a few mJy beam$^{-1}$. Some of the relative ease in calibrating LOFAR data at the present time is likely due to the low level of solar activity, meaning that ionospheric effects are rather benign. This situation is expected to change in the coming years, as the solar cycle reaches its peak. LOFAR imaging is even accessible at sub-arcsecond resolution, as shown by \cite{wucknitz_tv}. Our first efforts with LOFAR imaging involved fields containing only point sources. More recent efforts, highlighted in this contribution, show that LOFAR can also be used successfully to image extended sources.

The LOFAR imaging pipeline is the vehicle that will be used for automatically producing calibrated images immediately following the conclusion of an observing session. The fundamental components of the pipeline are in place. For many commissioning tests, the individual software components are used independently. Still, the pipeline has already been run end-to-end to image fields for which we have prior knowledge. Efforts are now underway to extend this functionality to arbitrary fields without human intervention.

One of the major improvements that will happen to LOFAR in the near future is the addition of station calibration, which maximizes the sensitivity of the individual stations and makes the station beams much more well behaved. Once the station calibration is in place, we will be able to produce images on a calibrated flux scale\footnote{This also relies on the availability of direction dependent corrections in the imaging stage; this feature is under development by the CImager team.}. Our ability to calibrate full polarization data (see \cite{beck_tv}) will also be significantly enhanced by the addition of station calibration.

\acknowledgments{The work described here relies on the efforts of many people who have been involved in the LOFAR project. We gratefully acknowledge the huge amount of work done by the entire LOFAR hardware and software development teams. We also thank the much larger group of commissioners who have taken part in other LOFAR Imaging Busy Weeks, beyond the Busy Week that produced the results that are the focus of this contribution. The full commissioning team has been instrumental in making the progress needed to successfully calibrate and image LOFAR data. The efforts of the commissioning team have also led to the development of the \emph{LOFAR Imaging Cookbook}, which is under constant development and is available for download from the following webpage: {\tt http://www.mpa-garching.mpg.de/$\sim$fdg/LOFAR\_cookbook/}}


\begin{thebibliography}{99}
  \bibitem{beck_tv} R.~Beck, \emph{Towards a New Era of Observing Cosmic Magnetic Fields}, in proceedings of \emph{ISKAF2010 Science Meeting}, \pos{PoS(ISKAF2010)003}.
  \bibitem{brentjens_2008} M.A.~Brentjens, \emph{Deep Westerbork observations of Abell 2256 at 350 MHz}, \emph{A\&A} {\bf 489} (2008) 69.
  \bibitem{debruyn_tv} A.G.~de Bruyn, \emph{Latest on imaging with LOFAR}, presented at \emph{ISKAF2010 Science Meeting}, {\tt http://www.astron.nl/sites/astron.nl/files/cms/PDF/10June11-LOFAR.pdf}
  \bibitem{clarke_ensslin_2006} T.E.~Clarke \& T.A.~En{\ss}lin, \emph{Deep 1.4 GHz Very Large Array Observations of the Radio Halo and Relic in Abell 2256}, \emph{AJ} {\bf 131} (2006) 2900.
  \bibitem{cohen_etal_2007} A.S.~Cohen et al., \emph{The VLA Low-Frequency Sky Survey}, \emph{AJ} {\bf 134} (2007) 1245.
  \bibitem{drury_1983} L.Oc.~Drury, \emph{An introduction to the theory of diffusive shock acceleration of energetic particles in tenuous plasmas}, \emph{RPPh} {\bf 46} (1983) 973.
  \bibitem{ensslin_etal_1998} T.A.~En{\ss}lin et al., \emph{Cluster radio relics as a tracer of shock waves of the large-scale structure formation}, \emph{A\&A} {\bf 332} (1998) 395. 
  \bibitem{falcke_etal_2007} H.D.~Falcke et al., \emph{A very brief description of LOFAR -- the Low Frequency Array}, \emph{HiA} {\bf 14} (2007) 386 [{\tt astro-ph/0610652}].
  \bibitem{hamaker_etal_1996} J.P.~Hamaker, J.D.~Bregman \& R.J.~Sault, \emph{Understanding radio polarimetry. I. Mathematical foundations}, \emph{A\&AS} {\bf 117} (1996) 137.
  \bibitem{intema_2009} H.T.~Intema, \emph{A sharp view on the low-frequency radio sky}, \emph{Ph.D. thesis} (2009).
  \bibitem{johnston_etal_2008} S.~Johnston et al., \emph{Science with ASKAP. The Australian square-kilometre-array pathfinder}, \emph{ExA} {\bf 22} (2008) 151.
  \bibitem{kale_dwarakanath_2010} R.~Kale \& K.S.~Dwarakanath, \emph{Spectral Index Studies of the Diffuse Radio Emission in Abell 2256: Implications for Merger Activity}, \emph{ApJ} {\bf 718} (2010) 939.
  \bibitem{kassim_etal_1993} N.E.~Kassim et al., \emph{Subarcminute resolution imaging of radio sources at 74 MHz with the Very Large Array}, \emph{AJ} {\bf 106} (1993) 2218.
  \bibitem{offringa_etal_2010} A.R.~Offringa et al., \emph{A LOFAR RFI detection pipeline and its first results}, in proceedings of {RFI mitigation workshop}, \pos{PoS(RFI2010)036}.
  \bibitem{rottgering_tv} H.J.A.~Rottgering, \emph{LOFAR and the low frequency Universe}, in proceedings of \emph{ISKAF2010 Science Meeting}, \pos{PoS(ISKAF2010)050}.
  \bibitem{swinbank_tv} J.~Swinbank, \emph{The LOFAR Transients Pipeline}, in proceedings of \emph{ISKAF2010 Science Meeting}, \pos{PoS(ISKAF2010)082}.
  \bibitem{vanweeren_etal_2009} R.J.~van Weeren et al., \emph{The discovery of diffuse steep spectrum sources in Abell 2256}, \emph{A\&A} {\bf 508} (2009) 1269.
  \bibitem{wucknitz_tv} O.~Wucknitz, \emph{Long baseline experiments with LOFAR}, in proceedings of \emph{ISKAF2010 Science Meeting}, \pos{PoS(ISKAF2010)058}.
\end{thebibliography}
\end{document}